# Who embraces AI in play? Exploratory modeling of player preference profiles toward game AI

Ting-Chen Hsu[1] *, *Student Member, IEEE*, Jiangxu Lin[1], Wenran Chen[1], Zheyuan Zhang[1], and Fei Qin[2] *
[1] *School of Animation and Digital Arts, Communication University of China, Beijing, China*
[2] *School of Information Engineering, Lanzhou City University, Lanzhou, China*

***Abstract*—Artificial intelligence is increasingly entering digital games through diverse functions. While prior work has shown that player attitudes toward game AI are strongly context-dependent, less is known about how these attitudes are structurally combined within different groups of players. This study addresses this gap by modeling players' cross-context AI acceptance as interpretable attitude profiles. Based on questionnaire data from 771 digital game players, we apply Archetypal Analysis (AA) to standardized acceptance ratings across eight representative AI application contexts in games. The analysis identifies seven distinctive profiles: AI-Skeptics, Broad AI-Supporters, Creative-Play Explorers, Experience-Oriented Supporters, Systemic Order Advocates, Emotion-Centered Supporters, and Governance-Skeptics. Exploratory one-vs-rest (OvR) logistic regressions further suggest that profile membership is associated with players' perceived AI literacy, gaming habits, disciplinary background, personality traits, and application-specific priorities. By shifting attention from isolated acceptance judgments to patterned preference structures, this study provides an exploratory empirical vocabulary for segmenting game AI audiences and offers preliminary design implications for more context-sensitive and player-sensitive AI integration in digital games.**

***Index Terms*—Human-AI interaction, digital games, player attitudes, typology, archetypal analysis.**

## I. INTRODUCTION

ARTIFICIAL intelligence (AI) is being embedded in multiple aspects of digital games, such as driving NPC dialogue [1], assist in storyline creation [2] and in-game content generation [3]. Prior work shows that players evaluate game AI differently across contexts, but it remains unclear whether these evaluations form interpretable cross-context player profiles [4], [5]. Existing studies have shown that evaluations of game AI vary across application contexts and intervention methods [6], and related work has examined how specific AI applications shape perceived player experience [7]. Yet less is known about whether players' evaluations across multiple AI applications form interpretable cross-context preference profiles, and whether these profiles are associated with individual player characteristics.

This exploratory study analyzes questionnaire data from 771 digital game players. Building on player typology research [8], we use Archetypal Analysis (AA) to identify

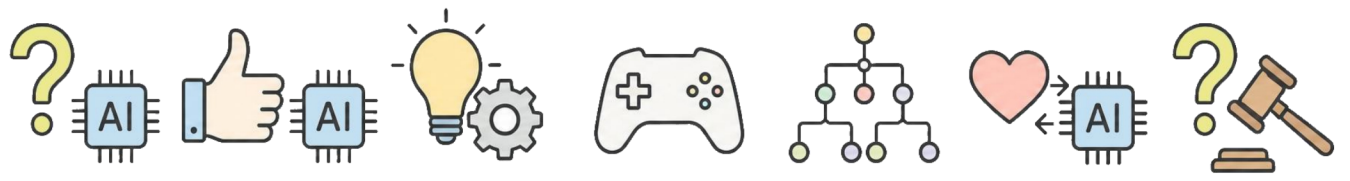

**Fig. 1.** This study uses Archetypal Analysis (AA) to identify seven exploratory profiles of players' acceptance preferences for AI in games.

extreme but interpretable profiles of game AI preference. This study mainly focuses on:

RQ1: What archetypal preference profiles can be identified from players' evaluations of different game AI application contexts?

RQ2: Which player characteristics are associated with players' dominant archetype or membership weights?

By shifting the unit of analysis from AI application contexts to player-level acceptance patterns, this study provides a preliminary framework for understanding heterogeneous player responses to game AI. The findings offer an empirical basis for the design, communication, and governance of AI functionalities tailored to diverse player profiles.

## II. METHODS

### *A. Measures*

A total of 814 responses were collected through an online questionnaire survey targeting digital game players recruited from Chinese social media platforms. The questionnaire contained self-designed items on demographic information, gaming habits, AI use, and attitudes toward AI applications in games. Perceived AI literacy was operationalized as a brief perceived AI familiarity index, calculated as the mean of two self-report items: participants' general frequency of AI use and their self-rated proficiency in using AI tools. Big Five personality traits were measured using the short Big Five inventory BFI-10 [9], with two items for each trait. To support content validity, we constructed eight game AI application contexts based on existing research [6]. Participants rated their acceptance of eight contexts, including driving intelligent NPCs, emergent narrative, dynamic balance adjustment, personalized recommendation, intelligent review and community governance, generating art assets, AI-supported co-creation gameplay, and gameplay self-evolution. All acceptance items were rated on a 5-point Likert scale. Before formal distribution, the questionnaire was pilot-tested with 50 participants, ensuring that participants could accurately understand all items and AI application contexts. The study has obtained informed consent from all participants and ethical approval from the author's institution. After excluding responses with abnormal completion times and straight-lined answers, 771 valid responses were retained for analysis.

*Corresponding authors. tingchenhsu.ac@gmail.com 15209446631@163.com
For any additional supplementary materials and raw data, please contact the corresponding authors.

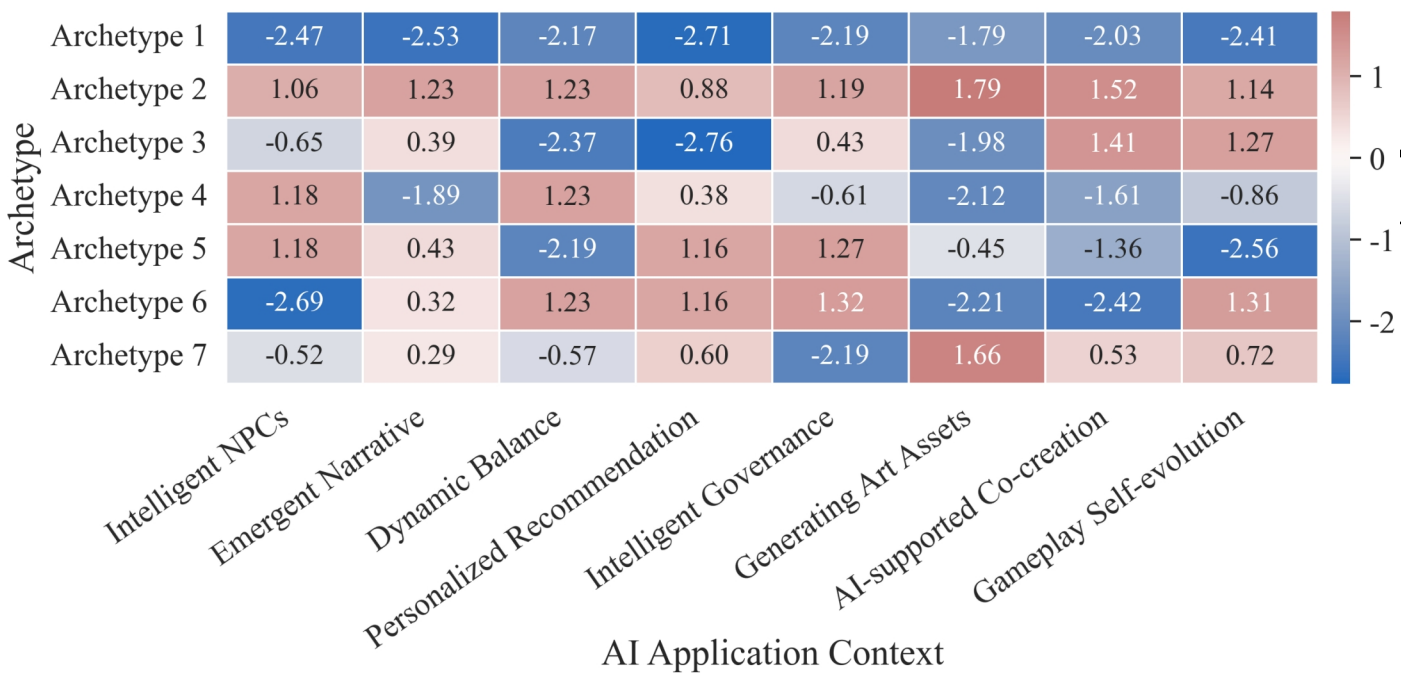

**Fig. 2.** Relative acceptance pattern of archetypes.

TABLE I
MODEL SELECTION DIAGNOSTICS FROM K=3 TO K=10

| k | RSS | MSE | Entropy | Mean MBC | Note |
|---|---|---|---|---|---|
| 3 | 3283.998 | 0.532 | 0.468 | 0.525 | Overly coarse |
| 4 | 2613.413 | 0.424 | 0.576 | 0.613 | Overly coarse |
| 5 | 2368.339 | 0.384 | 0.616 | 0.599 | Coarse |
| 6 | 2129.585 | 0.345 | 0.655 | 0.658 | Coarse |
| 7 | 1788.690 | 0.290 | 0.710 | 0.684 | Chosen |
| 8 | 1596.928 | 0.259 | 0.741 | 0.624 | Marginal type |
| 9 | 1540.809 | 0.250 | 0.750 | 0.629 | Marginal type |
| 10 | 1433.29 | 0.232 | 0.768 | 0.650 | Marginal type |

### *B. Data analysis*

To address RQ1, we used Archetypal Analysis (AA) to model players' relative acceptance patterns across the eight game-AI application contexts. Before model fitting, acceptance ratings across the eight AI application contexts were feature-wise standardized. The analysis therefore captures player-level acceptance patterns relative to the distribution of ratings within each application context. AA represents respondents as mixtures of extreme, interpretable orientations rather than discrete clusters (e.g., latent profile analysis).

AA models with k=3–10 were fitted and the final solution selected based on reconstruction error, interpretability, and bootstrap stability. Reconstruction error was assessed using the residual sum of squares (RSS), while interpretability was evaluated by whether each archetype showed a substantively meaningful pattern across the eight AI contexts. Bootstrap stability was examined by repeatedly resampling respondents, refitting the AA model, and matching the resulting archetypes to the reference solution using correlation-based optimal assignment. For interpretation, we examined the acceptance pattern of each archetype, the distribution of respondent membership weights, and the dominant archetype for each respondent. Profile-related attitude-reason variables were summarized descriptively to clarify the psychological rationale underlying each archetypal orientation.

To document model selection and reproducibility, we fitted AA solutions from k=3 to k=10. AA was implemented using the archetypes package in Python. For each k, we used farthest-point initialization with small random noise, 6 random initializations, a maximum of 80 iterations, and convergence tolerance 1e-4; the run with the lowest RSS was retained. Stability was assessed via 20 bootstrap resamples, refitting AA and matching archetypes using maximum profile correlation.

To address RQ2, one-vs-rest (OvR) logistic regression was used as an exploratory analysis to identify characteristics associated with respondents' dominant archetype. Predictors included demographics, gaming behavior, Big Five traits, and AI-related variables (literacy, usage, disclosure attention, openness). All multi-select items were one-hot encoded, and continuous predictors were z-standardized before model estimation. Models were estimated using maximum likelihood, reporting coefficients, odds ratios (OR), 95% confidence intervals, and p-values. Significance was set at $\alpha$=0.05.

## III. RESULTS

### *A. Descriptive statistics*

In the final sample, participants (N=771) had a mean age of 22.98 (SD=5.30), and included diverse gender identities (43.06% male, 52.66% female, 1.56% non-binary, 2.72% preferred not to disclose). Overall acceptance of AI in games was moderately positive but varied across application contexts, ranked from high to low as: personalized recommendation systems (3.82±0.99), driving intelligent NPC (3.82±0.92), dynamic balance adjustment (3.77±0.98), emergent narrative (3.72±1.08), intelligent review and community governance (3.68±1.13), gameplay self-evolution (3.56±1.07), AI-supported co-creation gameplay (3.48±1.56), and generating art assets (3.21±1.30). Importantly, the variation in overall acceptance levels indicates the presence of general pro-/anti-AI tendencies.

### *B. Archetypal AI preference profiles and interpretive labels*

Table I shows classification metrics ranging from k=3 to k=10, and the final result takes into account RSS, MSE, entropy, mean matched bootstrap correlations (MBC), and interpretability. The seven-archetype solution was selected as the most balanced model (Fig. 2). It introduced a substantively meaningful profile that remained merged in k=6, while also improving fit and bootstrap stability. Although k=8 further reduced error, its additional archetype is marginal and mainly split existing profiles, lowering stability and suggesting limited interpretive value. Labels were assigned based on each pattern and partial supporting attitude-reason summaries.

A1 (N=87, 11.28%) represents a consistently AI-skeptical profile, showing below-average acceptance across all contexts. We therefore label this profile "AI-Skeptics". In contrast, A2 is broadly AI-receptive and constitutes the largest dominant profile in the sample (N=328, 42.54%), showing above-average acceptance across all contexts. We label this profile "Broad AI-Supporters". These two archetypes anchor the global opposition between general resistance and general openness.

A3 (N=35, 4.54%) captures an exploratory creative-play profile: it is relatively open to co-creation and gameplay self-evolution, while rejecting dynamic balance, personalized recommendation, and AI asset generation. A3 also has the highest proportion of choosing "enhanced freshness" in the attitude variable, so we label this profile "Creative-Play Explorers". A4 (N=108, 14.01%) favors AI for system-facing adaptation, especially intelligent NPCs and dynamic balance, while rejecting more generative or production-oriented uses such as emergent narrative, AI-generated assets, and AI-supported co-creation. A4 also has a high proportion of choices with "enhanced experience" in the attitude variable results, thus we label it "Experience-Oriented Supporters". A5 (N=46, 5.97%) combines acceptance of recommendation, governance, and controlled AI assistance with rejection of dynamic balance and gameplay self-evolution, suggesting preference for managed rather than disruptive AI intervention. And given its high selection proportion in the attitude variable of "low stability", we label it "Systemic Order Advocates". A6 (N=25, 3.24%) similarly values governance and adaptive support, but is strongly negative toward intelligent NPCs, AI-generated assets and co-creation. The attitude variable results also show a clear tendency towards emotional skepticism in these three contexts, thus we label it "Emotion-Centered Supporters". A7 (N=142, 18.42%) shows a nearly opposite configuration: it is more accepting of AI-generated assets but strongly rejects intelligent governance. In the attitude-reason results, A7 also shows a broad "low stability" attitude towards intelligent governance, we therefore label this profile "Governance-Skeptics".

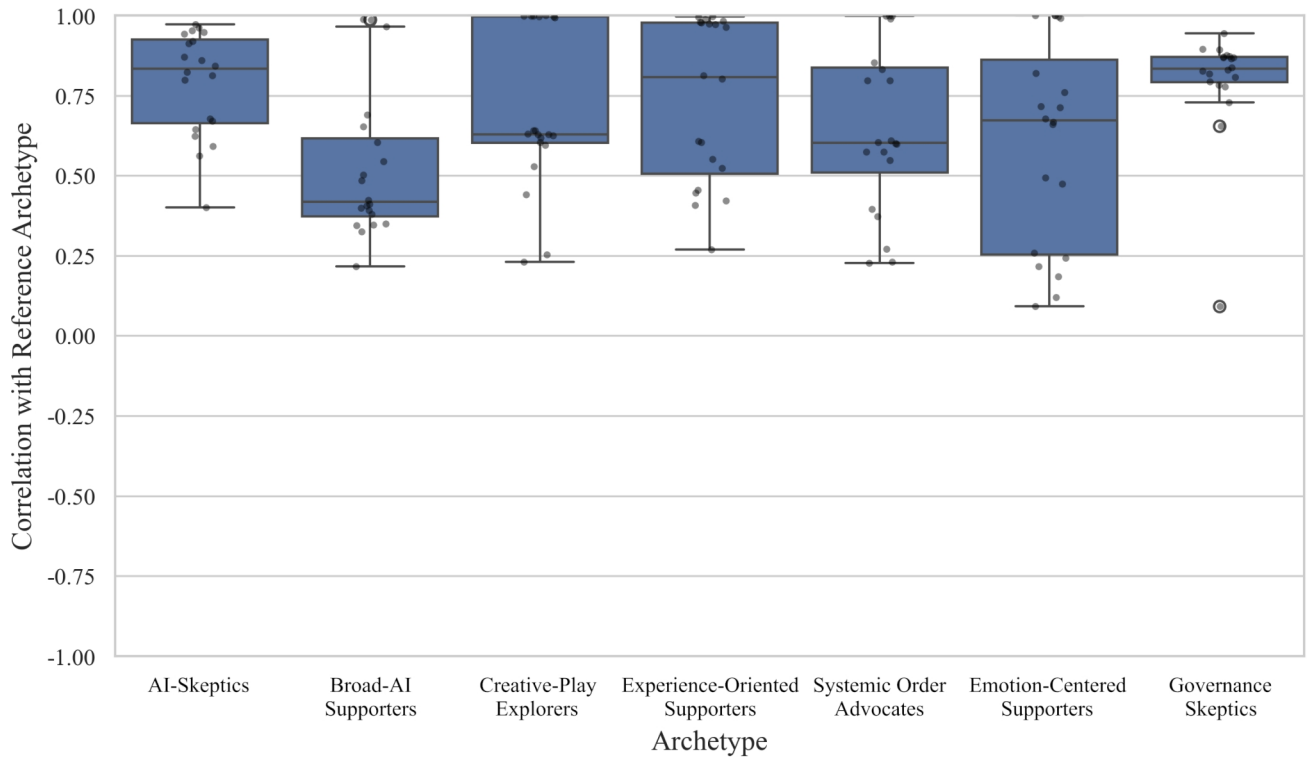

**Fig. 3.** Stability check results of MBC.

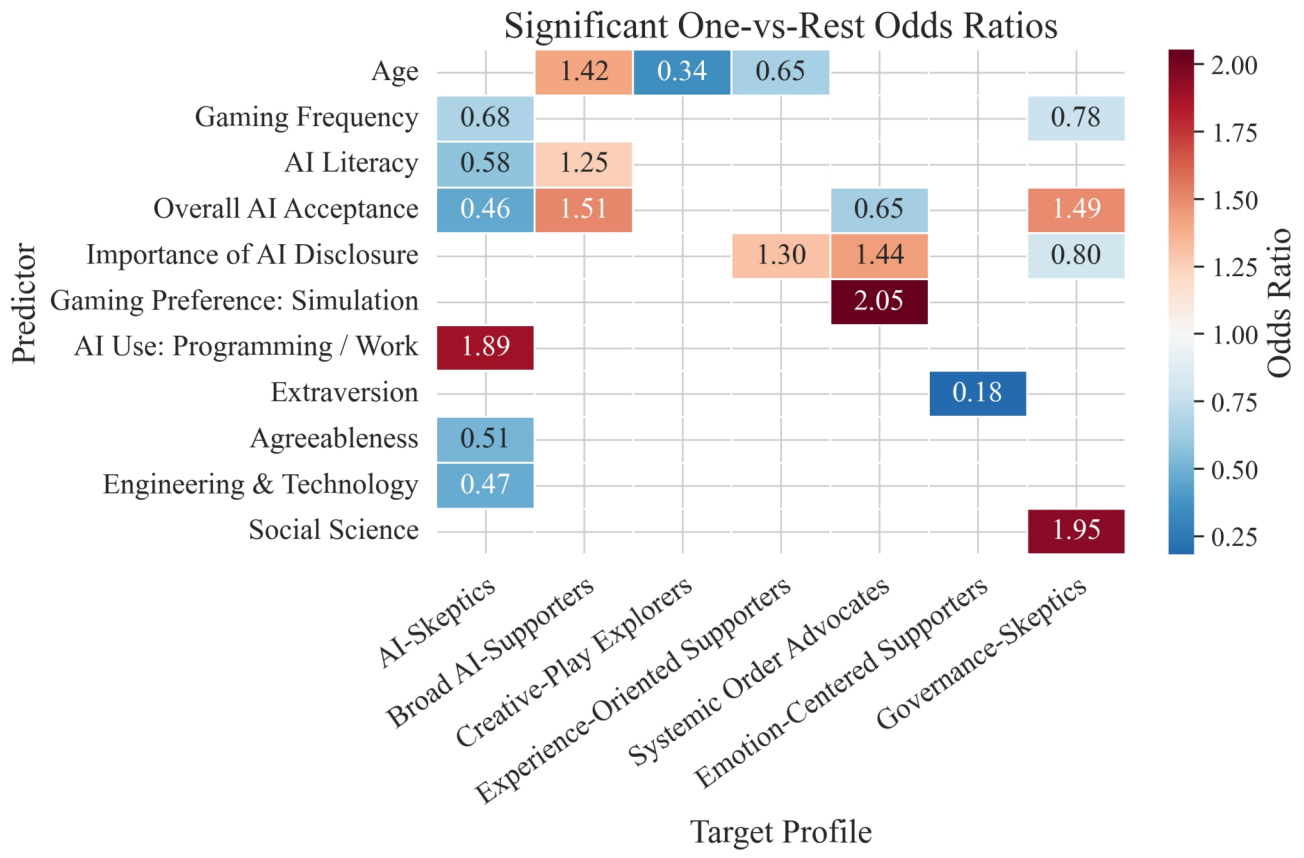

**Fig. 4.** Significant ($p < .05$) one-vs-rest odds ratios.

Bootstrap stability (Fig. 3) was uneven, with mean matched correlations from 0.521 (A2) to 0.795 (A7). A1, A4, and A7 were more distinctive, whereas A2 was diffuse and A3 weakly stable due to its small size. Therefore, the seven-profile solution should be interpreted as an exploratory map, with finer-grained profiles requiring validation.

TABLE II
MODEL PERFORMANCE OF EXPLORATORY ONE-VS-REST LOGISTIC REGRESSIONS

| Profile | Accuracy | AUC | F1 | Recall | McFadden $R^2$ |
|---|---|---|---|---|---|
| A1 | 0.748 | 0.798 | 0.316 | 0.529 | 0.256 |
| A2 | 0.548 | 0.613 | 0.539 | 0.621 | 0.109 |
| A3 | 0.594 | 0.387 | 0.031 | 0.143 | 0.103 |
| A4 | 0.632 | 0.548 | 0.174 | 0.273 | 0.121 |
| A5 | 0.690 | 0.545 | 0.111 | 0.333 | 0.124 |
| A6 | 0.748 | 0.664 | 0.093 | 0.400 | 0.192 |
| A7 | 0.594 | 0.706 | 0.442 | 0.862 | 0.076 |

### *C. Player characteristics associated with dominant profiles*

Because the OvR models were intended as exploratory follow-up analysis rather than confirmatory prediction models, we treat model performance and coefficients as descriptive indicators of potential associations. Overall, the OvR models provided moderate discrimination for A1 and A7, limited discrimination for A2, A4, A5, and A6, and very weak discrimination for A3. Full model-level indices are reported in Table II, while Fig. 4 visualizes predictors reaching $p < .05$.

The strongest contrast appears between global orientation toward AI and selective skepticism. Broad AI-Supporters (A2) were more likely to report higher overall AI acceptance (OR=1.51), perceived AI literacy (OR=1.25), and age (OR=1.42), suggesting a relatively mature and AI-familiar receptive group. In contrast, AI-Skeptics (A1) were less likely to report high AI literacy (OR=0.58), overall AI acceptance (OR=0.46), agreeableness (OR=0.51), and engineering background (OR=0.47), but were more likely to use AI for programming or work (OR=1.89). This pattern tentatively suggests that skepticism may not be reducible to lack of exposure alone, although this association should be interpreted as exploratory and requires validation in future samples.

Other profiles were differentiated by more specific motivations. Creative-Play Explorers (A3) had the lowest performance among all profiles (AUC=0.387), indicating that this small group was not reliably predictable from the related variables. Its only significant association was younger age (OR=0.34), consistent with a more exploratory orientation toward generative gameplay to a small extent. Experience-Oriented Supporters (A4) were linked to lower age (OR=0.65) and stronger perceived importance of AI disclosure (OR=1.30), indicating that support for immersive AI features may coexist with a demand for transparency. Systemic Order Advocates (A5) were less likely to show high overall AI acceptance (OR=0.65), but more likely to value AI disclosure (OR=1.44) and to prefer simulation games (OR=2.05), suggesting that this group may be especially attuned to rule-based, transparent, and controllable AI functions. Emotion-Centered Supporters (A6) were primarily distinguished by lower extraversion (OR=0.18), pointing to a possible affinity between less socially outgoing players and emotionally supportive AI functions. Governance-Skeptics (A7) showed higher overall

AI acceptance (OR=1.49) and social-science background (OR=1.95), but lower gaming frequency (OR=0.78) and lower emphasis on AI disclosure (OR=0.80). Together, these exploratory patterns suggest that player responses to game AI may be associated with combinations of perceived AI literacy, gaming habits, disciplinary background, and psychological traits, rather than a single pro- versus anti-AI dimension.

## IV. DISCUSSION

### *A. From isolated attitudes to player preference architectures*

Rather than treating AI applications independently, this study identifies recurring player-level acceptance architectures across functions. The seven archetypes reveal distinct configurations of design expectations, such as broad receptiveness, systemic caution, creative exploration, or governance skepticism. A2 further indicates that many players are broadly but diffusely tolerant of game AI. This shifts the analytical focus from whether a given AI function is acceptable to how different AI expectations cluster within players. In this sense, the archetypes offer a population-level map of game AI adoption: they show which forms of AI are likely to be endorsed together, which are separated by player values, and where potential tensions emerge between creativity, system control, and governance.

### *B. Design implications for segmented AI publics*

These results suggest that game AI should not be deployed as a uniform package for an imagined average player. The archetypes instead point to profile-sensitive design choices. For broadly receptive players, NPC dialogue, recommendation, and adaptive balancing may be bundled as an integrated AI layer, but with a visible global switch. For experience-oriented players, AI should be positioned as an experience-enhancing support layer, such as more responsive NPC behavior and smoother difficulty adaptation, rather than as a visible generator of narrative, assets, or co-created content. For creative-play explorers, AI may be better implemented as remixable co-creation tools or player-constrained generators rather than autonomous authorship. For governance-skeptical or generally skeptical players, moderation AI should provide explanation notices, appeal channels, and human-review fallback. Thus, the archetypes can inform concrete interface defaults, disclosure points, opt-in settings, and control granularity before game-AI deployment.

### *C. Limitations and future work*

This study has several limitations. First, the archetypes are derived from self-reported acceptance ratings, so they should not be interpreted as direct evidence of in-game behavior. Future work should test whether archetype membership predicts behavioral outcomes in live game environments. Second, the stability of the identified profiles is uneven. Future studies should validate the seven-profile structure with larger and more diverse samples, and examine whether fewer or more profiles emerge in specific game communities.

Third, several questionnaire items, including the game-AI acceptance items and the perceived AI literacy, were self-designed and not independently validated. They were kept brief to reduce respondents' cognitive burden across multiple AI contexts, and should be regarded as exploratory indicator rather than fully validated psychometric scales. Fourth, the China-based sample limits cultural generalizability, and future work should test whether similar archetypes emerge across different cultural, regional, and game-community contexts. Fifth, the OvR regression results should be interpreted as exploratory, hypothesis-generating associations. Future work should improve prediction by adding higher-dimensional behavioral variables, including generative-AI play experience, creative self-efficacy, and genre-specific histories, which may better capture small and ambiguous profiles such as A3. Finally, the present study captures a selected set of AI application contexts and player characteristics. As game AI systems evolve, more novel AI-mediated play, production, moderation, and monetization may activate different concerns. Future work should extend the model across more genres, platforms, and longitudinal exposure to AI systems.

## V. CONCLUSION

This study moves game-AI acceptance research from application-level analysis to player-level preference structures. Using AA on data from 771 players, we identified seven exploratory profiles that distinguish skepticism, broad support, creative-play support, experience-oriented support, systemic order concerns, emotion-centered support, and governance skepticism. These profiles offer a preliminary vocabulary for audience-sensitive AI design, suggesting that responsible game-AI integration should balance transparency, legitimacy, and control. Future work should examine how these archetypes generalize across game genres, cultural contexts, and more AI-mediated play in larger samples.